# EXCLUSIVE VECTOR MESON PRODUCTION AT HERA

A. Donnachie
Department of Physics and Astronomy, University of Manchester
Manchester M13 9PL, UK

P. V. Landshoff
University of Cambridge
Cambridge CB3 9EW, UK

## Abstract

Exclusive vector meson production from real and virtual photons is, for the most part, described well by soft pomeron exchange. However, there are some unsolved normalisation problems, and just a hint that an additional contribution from hard pomeron exchange may be needed.

It is well established that Regge theory [1] provides an economical and precise description of all total cross sections [2]. In particular, measurements of the total photoproduction cross section at HERA by the H1 [3] and ZEUS [4] collaborations are readily explained in terms of soft-pomeron exchange [2]. On the other hand, the rise of the the proton structure function at small $x$ certainly requires a different explanation, despite the expectation that the behaviour of $\nu W_2$ should be governed by Regge theory at small $x$ [5]. Further, recent data from ZEUS [6] on $\rho$ electroproduction appear to imply an energy dependence of the cross section, at fixed $Q^2$, much greater than would be expected from soft-pomeron dominance, despite the latter providing an excellent description of corresponding data at fixed target energies [7]. Elastic $\rho$ photoproduction, which has been measured recently by ZEUS [8], [9], in principle can provide additional information, and it is to this reaction we turn first.

Regge fits to the total cross sections are simply a sum of two powers:

$$\sigma_T = X s^\epsilon + Y s^{-\eta} \tag{1}$$

where the first term is identified as arising from pomeron exchange and the second from $\rho, \omega, f, a$ exchange. The effective powers $\epsilon$ and $\eta$ are given by [2]

$$\begin{aligned} \epsilon &\approx 0.08 \\ \eta &\approx 0.45 \end{aligned} \tag{2}$$

Application of Regge theory to hadronic elastic scattering is equally successful [10],[11]. The additional ingredients required for this are a knowledge of the form factors of the hadrons and the slopes of the $\rho, \omega, f, a$ and pomeron trajectories $\alpha_R(t)$ and $\alpha_P(t)$. Photoproduction of vector mesons provides an additional process, to which these ideas can be applied.

Extensive data on $\rho$, $\omega$ and $\phi$ photoproduction have been available for many years from fixed target experiments (see for example [12],[13],[14] and references quoted therein). At laboratory energies above 50 GeV or so, the $\rho$ and $\omega$ data are compatible with diffractive production i.e. pomeron exchange. At lower energies there is an increasingly strong contribution from non-pomeron $\rho, \omega, f, a$ exchange. For $\phi$ photoproduction, Zweig's rule has the consequence that the only exchange that couples both to the proton and the $\gamma - \phi$ transition vertex is the pomeron. The $\phi$ data are compatible with diffractive production over most of the energy range due to the resulting very small (perhaps zero) contribution from $\rho, \omega, f, a$ exchange.

The recent data on $\rho$ photoproduction from HERA [8],[9] greatly extend the lever arm and provide a more thorough test of the underlying ideas.

A calculation of diffractive $\rho$ photoproduction, with essentially no free parameters, can be obtained by assuming vector meson dominance [15], [16] and the additive quark model [17],[18]. The simplest version of vector meson dominance tells us that the forward cross section for $\gamma p \to \rho p$ should be given by

$$\frac{d\sigma}{dt}(t=0) = \alpha \, \frac{4\pi}{\gamma_\rho^2} \, \frac{d\sigma}{dt}(\rho^0 p \to \rho^0 p : t=0) \tag{3}$$

where $4\pi/\gamma_\rho^2$ is the $\rho$-photon coupling, which can be found from the $e^+e^-$ decay width of the $\rho$. Even if this simplest version of VMD is refined to include also the contribution from off-diagonal terms, for example $\rho' \to \rho$, vector dominance is not an exact science. So we shall use the simplest version.

The additive quark model assumes that single pomeron exchange couples to single quarks in hadrons. This cannot be true of double or multiple exchanges, and so the additivity cannot be exact. But from the accuracy of the model [2][19], it appears that single exchanges are dominant. The quark additivity does not seem to work nearly so well for $f$ and $a$ exchange, but we shall use it for want of any alternative. We consider $\rho^0 p \to \rho^0 p$, where the exchange can only be $C = +1$. According to the additive quark model, the forward amplitude is simply the average of the forward amplitudes for $\pi^+ p \to \pi^+ p$ and $\pi^- p \to \pi^- p$. By the optical theorem, $\sigma_T = \mathrm{Im} A(t = 0)/s$, and so the forward differential cross section is given by

$$\frac{d\sigma}{dt}(t = 0) = \frac{1}{16\pi}(1 + \rho^2)\sigma_T^2 \tag{4}$$

where $\rho$ is the ratio of the forward real part to the forward imaginary part of the scattering amplitude. Using the Donnachie and Landshoff [2] total cross section fits to $\pi^- p$ and $\pi^+ p$ scattering the total cross section for $\rho^0 p$ scattering is, in millibarns,

$$\sigma_T(\rho p) = 13.6 s^{0.08} + 31.8 s^{-0.45} \tag{5}$$

Because only $C = +1$ exchange contributes, the forward amplitude has the form

$$\begin{aligned} A &= A_P s^{0.08}(-\cos(\tfrac{1}{2}\pi\alpha_P(0)) + i\sin(\tfrac{1}{2}\pi\alpha_P(0))) \\ &+ A_R s^{-0.45}(-\cos(\tfrac{1}{2}\pi\alpha_R(0)) + i\sin(\tfrac{1}{2}\pi\alpha_R(0))) \end{aligned} \tag{6}$$

where $\alpha_P(0) = 1 + \epsilon$ and $\alpha_R(0) = 1 - \eta$, where

$$\begin{aligned} A_P \sin(\tfrac{1}{2}\pi\alpha_P(0)) &= 13.6 \\ A_R \sin(\tfrac{1}{2}\pi\alpha_R(0)) &= 31.8 \end{aligned} \tag{7}$$

so as to agree with (5). Thus

$$\begin{aligned} A_P &= 13.74 \\ A_R &= 41.95 \end{aligned} \tag{8}$$

and hence

$$A = 13.7((0.127s^{0.08} - 1.99s^{-0.45}) + i(0.992s^{0.08} + 2.31s^{-0.45})) \qquad (9)$$

In the narrow-width approximation the $e^+e^-$ decay width of the $\rho$ is given by

$$\Gamma_{\rho \to e^+e^-} = \frac{\alpha^2}{3} \frac{4\pi}{\gamma_\rho^2} m_\rho \qquad (10)$$

The $e^+e^-$ branching fraction of the $\rho$ is $(4.44 \pm 0.21)10^{-5}$, the full width is 151.5 MeV and the mass is 768.1 MeV [20]. These yield

$$\frac{4\pi}{\gamma_\rho^2} = 0.494 \pm 0.023 \qquad (11)$$

To avoid needing to make any assumptions about the slope of the differential cross section i.e. effectively assuming knowledge of the $\rho$ form factor, it is more convenient in the first instance to calculate the forward differential cross section rather than the total cross section. The former is frequently cited in experimental papers, and if not is readily found from the quoted total cross section and slope of the differential cross section. The energy dependence of the naive prediction as outlined above is correct, but the normalisation is too high. Multiplying by a factor of 0.84 provides an excellent description of all the data. This is illustrated in Fig.1.

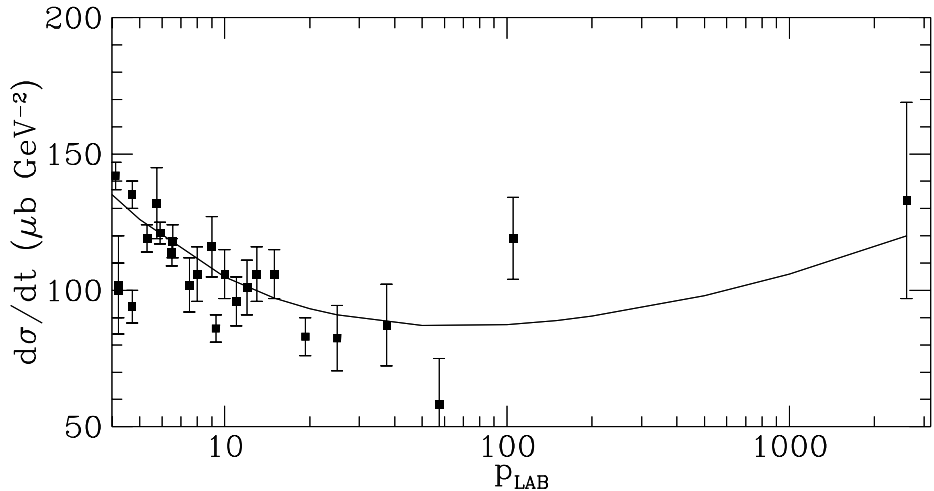

Figure 1: Data for the differential cross section for $\gamma p \to \rho p$ at $t = 0$, with soft-pomeron-exchange fit [2]

This normalisation difference does not represent a new problem: in fact it does not present any problem. It has been known for a long time that finite-width corrections to the $\rho \to e^+e^-$ decay rate are important: see for example Gounaris and Sakurai [21] and Renard [22]. In addition to this, there are the other intrinsic uncertainties in the vector dominance model, and the additive quark model is only accurate, in the pomeron dominated sector, to about 5%. Thus a factor of 0.84 between the naive prediction and the data is well within acceptable limits.

Wherever it can be experimentally checked the differential cross section for $\rho$ photoproduction is found to have the same slope at small $t$ as the $\pi p$ elastic cross section [12],[13]. Within the context of pomeron (and $\rho, \omega, f, a$) exchange, the obvious inference is that the form factor of the $\rho$ is the same as that of the pion. With this assumption, one can then see whether the energy dependence of the slope in $\rho$ photoproduction is compatible with that expected. In particular does the forward peak show the shrinkage expected at the higher energies where pomeron exchange dominates, and the forward slope is given by

$$b = b_0 + 2\alpha' \ln s \qquad (12)$$

where $b_0$ can be calculated from a knowledge of the $\rho$ and proton form factors. Unfortunately the data (not shown) are too poor to provide a decisive test. Although the predicted shrinkage of the forward peak, using the canonical value of $\alpha' = 0.25$, is compatible with the data up to HERA energies, a constant forward slope is not excluded.

In principle additional confirmation of the applicability of soft-pomeron phenomenology to vector meson photoproduction is provided by photoproduction of $\omega$, $\phi$, $J/\psi$. In practice, however, the data are either not of a quality or do not yet have the lever arm of HERA to provide a stringent test. At low energies $\omega$ photoproduction is complicated by a significant contribution from $\pi$-exchange. At energies sufficiently high for pomeron exchange to dominate there is no sensible difference between $\omega$ photoproduction and $\rho$ photoproduction, apart from the lower cross section due to the smaller $\omega$-photon coupling [12], [14]. For $\phi$ photoproduction, because of the pomeron dominance arising from Zweig's rule, the cross section should behave as $s^{2\epsilon}/b$ where $b$ is, as usual, the near-forward $t$-slope. Figure 2 shows a comparison with the data in the approximation that $b$ is a constant. The fit is not sensitive to letting the forward peak shrink in the canonical way i.e. by taking $b = b_0 + 2\alpha' \ln(s/s_0)$. A similar calculation for $\rho$, $\omega$ and $\phi$ elastic photoproduction has been performed by Schuler and Sjostrand [23].

The energy dependence is clearly compatible with present data. For the $J/\psi$, where again Zweig's rule applies, we simply note that away from the threshold region the fixed-target data are consistent with, but do not require, the standard $s^{2\epsilon}$ energy dependence. Data on $J/\psi$ photoproduction have recently become available from HERA [24] but in common with all other $J/\psi$ photoproduction data contain a significant contribution from events with single diffractive proton dissociation. Figure 3 shows the purely elastic curve [25], normalised to the low energy data. The discrepancy at high energies can be readily accounted for by a contribution from single proton dissociation which increases from about 2 nb at $\sqrt{(s)} = 10 GeV$ to about 12 nb at $\sqrt{(s)} = 100 GeV$ [24].

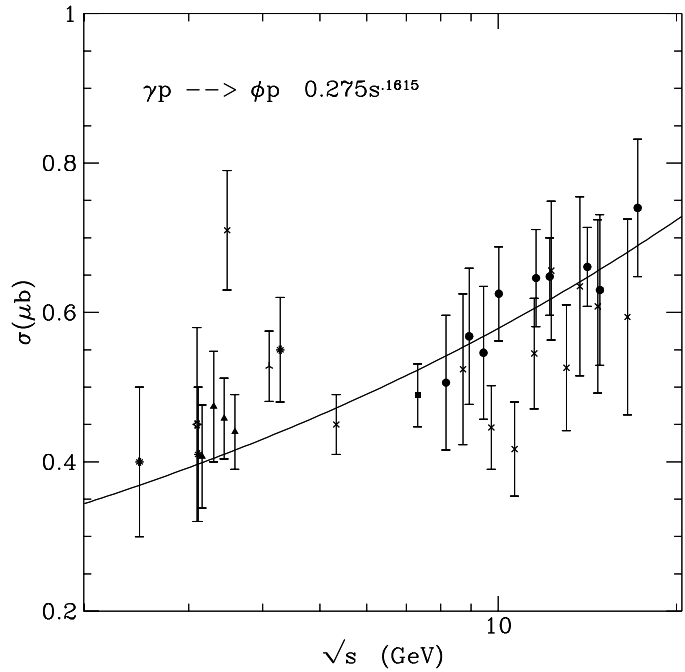

Figure 2: Data for $\gamma p \to \phi p$ with soft-pomeron-exchange fit [2]

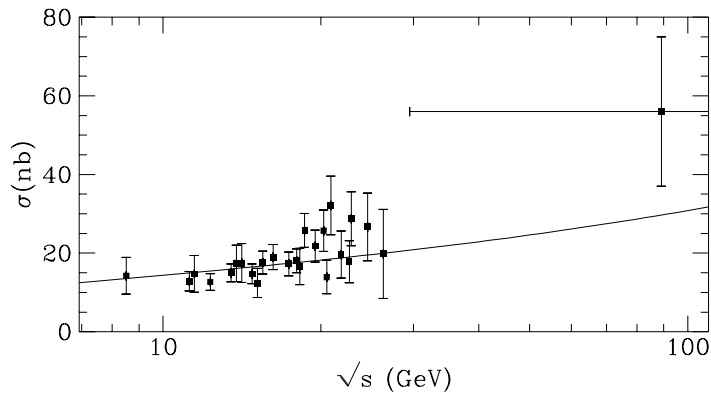

Figure 3: Data for $\gamma p \to J/\psi\, p$ with soft-pomeron-exchange fit [25]

Application of the same basic concepts to $\rho$ electroproduction [26] works remarkably well at fixed-target energies [27],[39] giving the correct normalisation, $Q^2$-dependence, $t$-dependence and $\rho$ alignment. This is seen from the upper curve and data points in Fig.4. The lower data points in the figure are for elastic $\phi$ electroproduction. The corresponding curve is calculated in the same way as for elastic $\rho$ electroproduction, but with a correction to the overall normalisation. The correction factor is 0.53 and is arrived at by using the additive quark model to calculate the cross section for strange-quark scattering off a proton from the total cross sections for $Kp$ and $\pi p$ scattering. The pomeron-exchange contribution to the $K^{\pm}p$ total cross section is [2] 87% as large as for the $\pi^{\pm}p$ total cross section. Thus the effective coupling of the pomeron to the strange quark is weaker than to the light quarks, only 73%. The square of this then appears in $\phi$ elastic photoproduction and electroproduction.

It is interesting that the absolute normalisation for both $\rho$ and $\phi$ production by virtual photons is not quite given correctly by the model. We have seen earlier that for $\rho$ photoproduction the calculation had to be renormalized downward by a factor of 0.84, which is within the uncertainty limits of vector dominance. For $\phi$ photoproduction the required factor is close to 1/2 [14],[29], which clearly implies a breakdown of simple vector dominance. Part of this factor can be explained by $\omega - \phi$ mixing through the $3\pi$ channel, a possibility first discussed by Ross and Stodolsky [30]. Some simple quark-model ideas developed by Gerasimov [31] and by Bauer and Yennie [32] suggested a small negative mixing amplitude, the existence of which is confirmed experimentally. The problem is well described in the reviews of Bauer et al [33] and of Leith [34]. However this explanation is not sufficient as the mixing decreases the cross section only by about 12%. There are two possible explanations for this, and probably both are involved. One is that the vector dominance becomes less reliable as the mass of the vector meson increases. Another is that the problem is due to specific wave-function effects, which are associated with the $\phi$ having comparatively small radius [35][36] and which should disappear at large $Q^2$. We shall comment more explicitly on this below. We are not persuaded that shadowing corrections are of significant magnitude [37]; our reasons are spelled out in reference [2]

The explanation of the magnitude of $\phi$ elastic electroproduction has interesting consequences for $J/\psi$ elastic electroproduction. The uncorrected calculation of [26] predicts that $J/\psi$ elastic electroproduction is actually greater than $\rho$ elastic electroproduction at large $Q^2$. If we simply multiply this by the apparent ratio of $J/\psi$ to $\rho$ elastic photoproduction, obtained by assuming that vector dominance may still be used for the $J/\psi$, we have a suppression of two orders of magnitude. On the other hand, if we believe that most of the suppression of $J/\psi$ elastic photoproduction is due to the breakdown of vector dominance and/or wave-function effects, with only some small contribution from a reduction of the strength of the pomeron coupling to the charm quark, then the $J/\psi$ elastic electroproduction cross section will be comparable to that of the $\rho$. This latter hypothesis does seem to be in accord with the EMC data [38] which agree broadly with the calculation at large $Q^2$. However as those data surely have some contamination from proton dissociation there is room for some suppression of the coupling of the pomeron to the charm quark, but the suppression cannot be large. This in turn implies that the cross section for charm scattering is rather larger than is usually assumed from analysis of the $J/\psi$ photoproduction data.

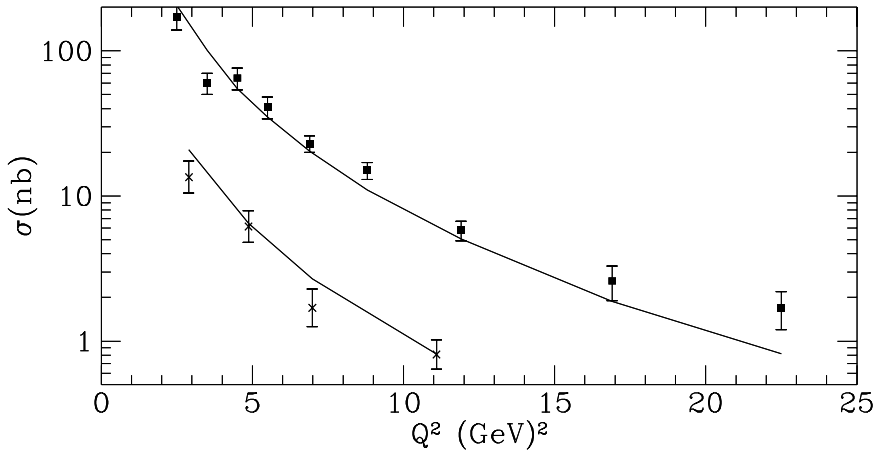

Figure 4: NMC data [39] for $\gamma^* p \to \rho p$ with prediction from reference [26]

For $\rho$ elastic electroproduction, comparison of the NMC data with the initial data from HERA [6] indicates that the energy dependence, at fixed $Q^2$, is very much stronger than $s^{2\epsilon}$ if $\epsilon \sim 0.08$. We shall assume that this is a real effect, even though past history has shown that [28][39] that there are very real experimental difficulties in removing from elastic electroproduction the contamination from events where the proton breaks up.

To explore this further requires a deeper look at the pomeron itself. The soft pomeron is non-perturbative in origin and is almost certainly a consequence of non-perturbative gluon exchange [35]. A simple model of the exchange of two non-perturbative gluons is compatible with much of pomeron phenomenology. In particular it can be applied to $\rho$ electroproduction [40],[28]. The relevant diagrams are those of Fig.5a which are applicable for $0 \le t \le 1$ GeV$^2$. The two diagrams tend to cancel at large $Q^2$ and together give a factor $\sim 1/Q^2$ in the amplitude and hence lead, for $\gamma^* p \to \rho^0 p$, to $d\sigma/dt \sim 1/Q^6$ when $Q^2 \ge 8$ GeV$^2$. Both the two-gluon exchange model and the phenomenological pomeron model predict that longitudinal polarisation should dominate at large $Q^2$, which is confirmed by experiment. As for the phenomenological pomeron, the energy dependence $s^{2\epsilon}$ has to be put in by hand. A key question, which is highlighted by the preliminary ZEUS data [6], is whether at large $Q^2$ the energy dependence still requires $\epsilon \approx 0.08$, corresponding to soft pomeron exchange, or whether the effective power is now greater. If it is, this would naturally be interpreted as there being also a contribution from hard pomeron exchange[41]. One test of whether it is still soft pomeron exchange that is at work is whether the $t$-dependence is correctly predicted: one would not expect the hard pomeron to give the same exponential slope. Remember, however, that any contamination from events in which the proton dissociates will have a marked effect on the measured slope.

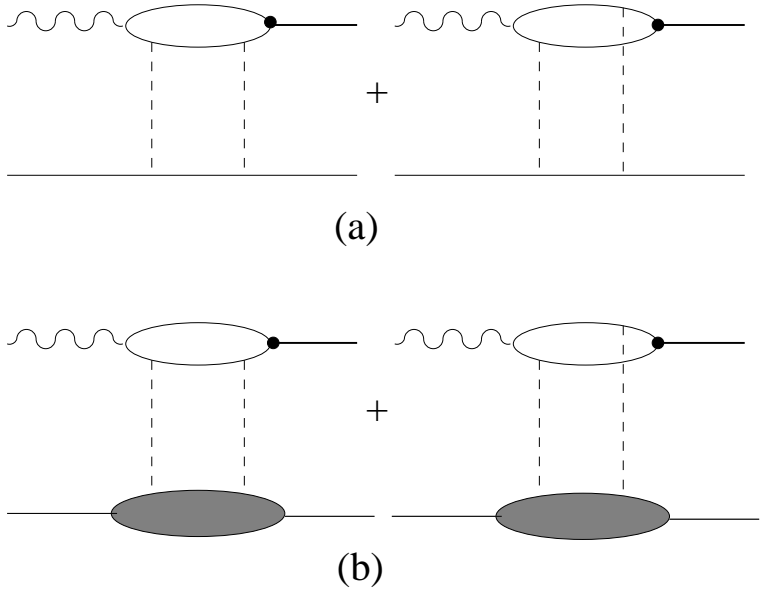

Figure 5: (a) simple model for $\gamma^* p \to \rho p$ and (b) refinement where the simple lower structure is replaced with the complete gluon structure function of the proton

The two-gluon exchange mechanism provides an insight into possible wave function effects. If the two gluons couple to different quarks in a hadron, then the loop momentum has to pass through the hadron wave function. Landshoff and Nachtmann [35] have shown that the contribution of this process, relative to that where the two gluons couple to a single quark, is $o(a^2/R^2)$ where $a$ is the correlation length of the gluon condensate in the vacuum and $R$ is the radius of the hadron. This term comes in with the opposite sign to that in which the two gluons couple to the same quark and so there be a partial cancellation. Although there is some uncertainty as to the precise value of $a$ it is certainly appreciably smaller than the value of $R$ for hadrons with light valence quarks. Thus the contribution from diagrams in which the two gluons couple to different quarks is small, in conformity with standard pomeron phenomenology. However the value of $R$ for hadrons consisting of heavy quarks, e.g. $J/\psi$, is much more comparable to $a$ and there is considerable cancellation between the contributions from the two different diagrams. This is fully accounted for in the two-gluon exchange model of vector-meson electroproduction, and there is no need for anything beyond the smaller nonpertubative coupling of a gluon to a heavy quark.

An interesting attempt [42][43] to include the energy dependence naturally relates the amplitude at zero momentum transfer $\Delta$ to the gluon structure function of the proton. The diagrams of Fig.5a are replaced by those of Fig.5b. Just as in the simpler model, the production of longitudinally polarized $\rho$ mesons will dominate at high $Q^2$, from longitudinally polarized virtual photons, yielding a $1/Q^6$ dependence:

$$\frac{d\sigma}{dt}\Big|_{\Delta=0} = \frac{A}{Q^6}(\alpha_s(Q^2)\,xg(x,Q^2))^2 \tag{13}$$

where $g(x,Q^2)$ is the gluon density of the proton and $x = Q^2/2\nu$. If the gluon structure function rises faster than $x^{-0.08}$ at small $x$ (which is the implication of the initial HERA results on $F_2$) then the $\rho$ electroproduction will equally rise faster. In the formalism of [43] the constant $A$ of eqn(13) depends on parameters of the $\rho$ wave function, and it is possible to choose these within sensible limits [4] to get agreement with the initial HERA data. However, there is a big difficulty: the connection (13) with the gluon structure function requires an extrapolation of the data to zero momentum transfer **vector** $\Delta$, where $t = \Delta^2$. Such an extrapolation is impossible, as it would require a large variation in the mass of the $\rho$ so as to reach $m_\rho^2 = -Q^2$. Simply extrapolating to $t = 0$ keeping $m_\rho$ fixed does not achieve $\Delta = 0$; it only makes $\Delta$ lightlike, and makes the difference of the $x$-parameters of the two gluons equal to $Q^2/2\nu$, rather than each of them taking this value. Sadly, therefore, we believe that the formula (13), involving the gluon structure function is not correct, beyond perhaps in a very approximate way.

*This research is supported in part by the EU Programme "Human Capital and Mobility", Network "Physics at High Energy Colliders", contract CHRX-CT93-0357 (DG 12 COMA), and by PPARC. We are pleased to acknowledge discussions with Jean-René Cudell and Markus Diehl.*